\documentclass[published]{JHEP3}
\JHEP{00(2009)000}

\usepackage{amsmath}

\def\beq{\begin{equation}}
\def\eeq{\end{equation}}
\def\bea{\begin{eqnarray}}
\def\eea{\end{eqnarray}}

\title{Non-Pauli Effects from Noncommutative Spacetimes}

\author{A. P. Balachandran \\ Department of Physics, Syracuse University, Syracuse,
NY 13244-1130, USA \\ The Institute of Mathematical Sciences, CIT
Campus, Taramani, Chennai 600 113, India \\ E-mail:
\email{bal@phy.syr.edu}}

\author{Pramod Padmanabhan \\ Department of Physics, Syracuse University, Syracuse, NY
13244-1130, USA \\ The Institute of Mathematical Sciences, CIT
Campus, Taramani, Chennai 600 113, India  \\ E-mail:
\email{ppadmana@syr.edu}}

\received{\today} \accepted{\today}

\preprint{SU-4252-907}

\abstract{Noncommutative spacetimes lead to nonlocal quantum field
theories (qft's) where spin-statistics theorems cannot be proved.
For this reason, and also backed by detailed arguments, it has been
suggested that they get corrected on such spacetimes leading to
small violations of the Pauli principle. In a recent
paper~\cite{Pauli}, Pauli-forbidden transitions from spacetime
noncommutativity were calculated and confronted with experiments.
Here we give details of the computation missing from this paper. The
latter was based on a spacetime $\mathcal{B}_{\chi\vec{n}}$
different from the Moyal plane. We argue that it quantizes time in
units of $\chi$. Energy is then conserved only mod
$\frac{2\pi}{\chi}$. Issues related to superselection rules raised
by non-Pauli effects are also discussed in a preliminary manner.}


\begin{document}
 The spin-statistics theorem in three or more
dimensions has been proved in many ways in local relativistic qft's.
It assumes its comprehensive form in the work of Doplicher and
Roberts~\cite{DR, DHR}. It states that identical tensorial particles
are bosons and identical spinorial particles are fermions. The
proofs of this theorem require the axioms of local relativistic
qft's. Deep extensions of the theorem to qft's on gravitational
backgrounds exist~\cite{Fredenhagen, Verch}, but they too require
spacetime commutativity and a form of locality.

 It is reasonable to expect that the spin-statistics connection
and its emergent physics can get modified in models where spacetime
commutativity and locality do not hold. We made a suggestion along
these lines for qft's on the Moyal plane~\cite{ApbGM}. A subsequent
paper by Chakraborty et al~\cite{Biswajit} developed this idea and
showed in a striking calculation that the Pauli repulsion between
fermions, infinite for zero separation on commutative spacetimes,
softens to a finite value on the Moyal plane. Applications of this
effect to statistical mechanics, superconductivity, and
Chandrasekhar limit either exist or are in progress~\cite{Sachin's}.

But these papers do not explicitly consider Pauli-forbidden
transitions.

With precision experiments at increasingly shorter length and time
scales, it is now timely to question principles of local qft's such
as Lorentz invariance, $CPT$ theorem and the spin-statistics
connection. As regards the last, there exist excellent experiments
on Pauli-forbidden transitions, but there is a scarcity of good
models to confront data, those of Greenberg and coworkers being
among the exceptions. These are reported or reviewed in~\cite{Milan,
Trieste, Greenberg} where also much existing information is
surveyed. A desirable model will have a small parameter $\chi$,
$\chi=0$ giving back the standard treatment. Here we develop such an
approach adapted to treat Pauli-violating atomic and nuclear
transitions.

Our model is based on a spacetime $\mathcal{B}_{\chi\vec{n}}$
different from the Moyal plane $\mathcal{A}_{\theta}$. The latter
also seems to predict the exotic effects we look for, but the
calculations get complicated. Just as in the case of
$\mathcal{A}_{\theta}$, $\mathcal{B}_{\chi\vec{n}}$ too can be
described in terms of a Drinfel'd twist element $F_{\chi\vec{n}}$(defined in
Eq.({\ref{twist})). So the Poincar\'{e} group algebra $\mathbb{C}\mathcal{P}$ can act on
$\mathcal{B}_{\chi\vec{n}}$ as a Hopf algebra if its coproduct is
deformed. Compatibility with this action requires that we deform the
standard symmetrization or flip operator $\tau_0$ to
\beq{\tau_{\chi\vec{n}}=F^{-1}_{\chi\vec{n}}\tau_0F_{\chi\vec{n}}.}\eeq
That changes the symmetrization and anti-symmetrization of wave
functions and leads to novel physics. The details we need about the
modified flip $\tau_{\chi\vec{n}}$ and the deformed Hopf algebra of
$\mathbb{C}\mathcal{P}$ are in Sec.1.

A typical Pauli-forbidden transition can occur in neutral beryllium
with two electrons in the ground state and the remaining two
electrons in the excited state: the transition of the excited
electrons to the ground state is Pauli-forbidden on the commutative
spacetime $\mathcal{B}_0$. But it occurs on
$\mathcal{B}_{\chi\vec{n}}$ and we calculate its rate. It involves
new physics, relying on the fact that the direction of the unit
vector $\vec{n}$ effectively changes with earth's rotation and
movements. These are very swift events for noncommutative
corrections induced by $\chi$, so that the sudden approximation is
appropriate to treat $\chi$-dependent atomic or nuclear
phenomena.(We do not consider TeV scale gravity~\cite{RSund}.)

 When $\vec{n}$ changes to $\vec{m}$ by earth's fast motions, twisted
fermions with $\tau_{\chi\vec{n}}=-1$ in the sudden approximation
become superpositions of both twisted fermions and twisted bosons
($\tau_{\chi\vec{m}}=\mp 1$) leading to the above process.

We are looking for violations of transitions which are strictly forbidden
in standard quantum theory. The calculations are greatly simplified if 
spin-orbit coupling is neglected. The latter will of course give 
corrections to our final answer. But in this paper, our focus is on 
establishing that such violations exist and {\it estimating} their magnitudes.
The corrections due to spin-orbit coupling will not rule out these violations
or significantly affect these estimates. For these reasons, we will ignore 
spin-orbit coupling.

Noncommutative spacetimes emerge from quantum gravity and Planck-
scale physics. Thus we are using atomic and nuclear phenomena to probe very
high energy physics. Our results are not expected to have much bearing
on low energy phenomena. 

Section 2 describes the two-electron energy eigenstates on
$\mathcal{B}_{\chi\vec{n}}$.

In section 3, we calculate what becomes of these state vectors when
$\vec{n}$ rapidly changes to $\vec{m}$. We explicitly find the
twisted Bose components induced in certain twisted Fermi levels of
$\mathcal{B}_{\chi\vec{n}}$. This enables us to calculate the rate
$R$ of transition of the excited electrons to the fully occupied
ground level for a sufficiently generic perturbation. $R$ depends on
$\vec{n}.\vec{m}$, but since $\vec{m}$ and $\vec{n}$ keep changing,
we average them to get an average rate $\langle R\rangle$.

Comparison with experiments are best done by developing a formula
for a branching ratio $B$ where the effects not specific to
noncommutativity may largely cancel. So we divide $\langle R\rangle$
by a typical rate for an allowed atomic or nuclear transition and
find a $B$. It is $O((\chi\Delta E)^2) $ where $\Delta E$ is a
suitable energy difference.

The expression for $B$ and the available atomic and nuclear
experiments give bounds on $\chi$. The use of $B$ away from its
original context is justified as remarked above, $B$ being a ratio.
In any case, our bounds are rough. They are reported in Sec.4. The
best ones come from neutrino signals of forbidden
processes~\cite{Borexino, SKamiokande, VIP} and give $\chi\gtrsim
10^{24}\textrm{TeV}$. This does seem an excessively stringent bound
suggesting further checks on its validity. As it stands, it suggests
an energy scale beyond Planck scale.

The focus of section 5 shifts away from Pauli principle and probes
other features of $\mathcal{B}_{\chi\vec{n}}$. We show that time
translation gets quantized on $\mathcal{B}_{\chi\vec{n}}$ in units
of $\chi$. Elsewhere this effect has been discussed in
detail~\cite{Baltrgpaulo, Balpaulo, Chaitime} and it has been proved
that energy is conserved only mod $\frac{2\pi}{\chi}$ in scattering
processes. A formal scattering theory has also been developed. Thus
$\mathcal{B}_{\chi\vec{n}}$ predicts much new physics. Its potential
applications to higher dimensional models is also pointed out in
section 6.

In the final section 6, we briefly consider the Moyal plane
$\mathcal{A}_{\theta}$ and argue that Pauli-forbidden transitions
exist there as well although computations become more involved. We
also comment on superselection rules and how they get
violated in our models. Their description using qft's is commented
on as well.

\section{The Spacetime $\mathcal{B}_{\chi\vec{n}}$}

The elements of  $\mathcal{B}_{\chi\vec{n}}$ are functions on the
Minkowski space $M^4$. If $x_{\mu}$ are coordinate functions
transforming under the Poincar\'e group $\mathcal{P}$ in the
standard manner, the algebra $\mathcal{B}_{\chi\vec{n}}$ is
characterized by the relations
\beq\label{Bplane}{[x_0,x_i]=i\chi\epsilon_{kij}n_kx_j,}\eeq
\beq{[x_i,x_j]=0,~~~~i,j=1,2,3}\eeq where $x_0$ is the time function
and $\vec{n}$ is a fixed three-dimensional unit vector.

A product map $m_{\chi\vec{n}}$ of two functions $f$, $g$, which
leads to Eq.(\ref{Bplane}) is given by
\beq\label{multi}{m_{\chi\vec{n}}(f\otimes g) =
fe^{\frac{1}{2}\chi(\overleftarrow{\partial_t}\vec{n}\cdot\vec{L}-
\vec{n}\cdot\overleftarrow{L}\vec{\partial_t})}g}\eeq where $\vec{L}=
-i\chi \vec{x}\wedge \vec{\nabla}$ is orbital angular momentum and
generates rotations. The product in Eq.(\ref{multi}), is associative
since $[\partial_t, \vec{n}\cdot\vec{L}]=0$. Equation (\ref{multi})
defines $\mathcal{B}_{\chi\vec{n}}$.

We can write Eq.(\ref{multi}) in terms of the twist element
\beq\label{twist}{F_{\chi\vec{n}}=e^{\frac{1}{2}\chi(\partial_t\otimes\vec{n}\cdot\vec{L}-\vec{n}\cdot\vec{L}\otimes\partial_t)}}\eeq
as follows: \beq{m_{\chi\vec{n}}= m_0\cdot F_{\chi\vec{n}},}\eeq
\beq{m_{\chi\vec{n}}(f\otimes g)=m_0[F_{\chi\vec{n}} f\otimes
g]}\eeq where $m_0$ is point-wise multiplication : \beq{m_0(f\otimes
g)(p) = f(p)g(p), ~~p = \textrm{a point of}~M^4.}\eeq

The algebra $\mathcal{B}_{\chi\vec{n}}$ is well-suited for deforming
dynamics with spherical symmetry as in atomic physics with its
central potentials. For the same reason, it is well-adapted to
deform quantum fields on black hole backgrounds. The Moyal plane is
awkward to deal with in either case (See however~\cite{Sachin's}).

The form of the twist element in Eq.(\ref{twist}) in a generic representation
carrying the action of $\mathbb{C}\mathcal{P}$ is known from the general theory
of Hopf algebras~\cite{Chaichian, Aschieri}. Thus  
in a generic representation carrying the action of
$\mathbb{C}\mathcal{P}$, $\vec{L}$ becomes the rotation generator
$\vec{J}$ and $i\partial_t$ the translation generator $P_0$. If
$G_{\chi\vec{n}}$ is the generic form of $F_{\chi\vec{n}}$, then
\beq\label{gtwist}{G_{\chi\vec{n}}=e^{-\frac{i}{2}\chi(P_0\otimes\vec{n}\cdot\vec{J}
- \vec{n}\cdot\vec{J}\otimes P_0)}.}\eeq We note that this form of $G_{\chi\vec{n}}$
is correct for any model which has rotation and time translation symmetry.
Relativistic invariance is not called for.

Drinfel'd's original work~\cite{Drinfeld} and subsequent
developments by Aschieri et al.~\cite{Aschieri} and Chaichian et
al.~\cite{Chaichian} show that $\mathbb{C}\mathcal{P}$ acts as a
Hopf algebra $H\mathcal{P}$ if its coproduct is modified by the
Drinfel'd twist $G_{\chi\vec{n}}$ to $\Delta_{\chi\vec{n}}$:
\beq{\Delta_{\chi\vec{n}}(g) := G_{\chi\vec{n}}^{-1}(g\otimes
g)G_{\chi\vec{n}},~~~~g\in\mathcal{P}.}\eeq

 For $\chi=0$, when noncommutativity is absent, symmetrization and
anti-symmetrization is achieved using the projectors
$\frac{1\pm\tau_0}{2}$. $\tau_0$ here is the flip operator: if
$\mathcal{H}$ is a Hilbert space carrying a representation of
$\mathcal{P}$ or one of its subgroups, and $\alpha$, $\beta\in
\mathcal{H}$, $\tau_0(\alpha\otimes\beta)=\beta\otimes\alpha$. This
flip commutes with $\Delta_0(g)$ and is Poincar\'e invariant for
$\chi=0$.

But for $\chi\neq 0$, \beq{\tau_0 G_{\chi\vec{n}}=
G_{\chi\vec{n}}^{-1} \tau_0}\eeq and $\tau_0$ fails to commute with
$\Delta_{\chi\vec{n}}(g)$: the projectors $\frac{1\pm\tau_0}{2}$ are
not Poincar\'e invariant for $\chi\neq 0$. Hence we must deform
$\tau_0$ suitably. Such a deformed flip operator is the twisted flip
operator
\beq{\tau_{\chi\vec{n}}=G_{\chi\vec{n}}^{-1}\tau_0G_{\chi\vec{n}}=G_{\chi\vec{n}}^{-2}\tau_0,~
\tau_{\chi\vec{n}}^2=1.}\eeq Thus if $\mathcal{H}$ is a
representation space for $\mathbb{C}\mathcal{P}$ or one of its
generic subgroups, and
$\alpha\otimes\beta\in\mathcal{H}\otimes\mathcal{H}$, the twisted
bosons and fermions are images of $\mathcal{H}\otimes\mathcal{H}$
under the projectors $\frac{\mathbb{I}\pm\tau_{\chi\vec{n}}}{2}$:
\beq\label{Twisted Bosons}{\textrm{Twisted Bosons:}~
\mathcal{H}\otimes_{S_{\chi\vec{n}}}\mathcal{H} :=
\frac{1+\tau_{\chi\vec{n}}}{2}\mathcal{H}\otimes\mathcal{H}}\eeq
\beq\label{Twisted Fermions}{\textrm{Twisted Fermions:}~
\mathcal{H}\otimes_{A_{\chi\vec{n}}}\mathcal{H} :=
\frac{1-\tau_{\chi\vec{n}}}{2}\mathcal{H}\otimes\mathcal{H}.}\eeq
The full justification of Eq.(\ref{Twisted Bosons}) and Eq.(\ref{Twisted Fermions})
will take us too far into Hopf algebra theory and material which has been
extensively treated elsewhere. [See for example~\cite{ApbGM}.]

 We note that $\mathcal{H}$ can be the Hilbert space of an electron
with spin in the central potential of a nucleus. The single particle
symmetry group $G$ we then focus on is $SU(2)\times\mathbb{R}$ where
$SU(2)$ is the (two-fold cover of the) rotation group acting also on
spin and rotating around the nuclear center, and $\mathbb{R}$ is the
time translation group. The generator $P_0$ of $\mathbb{R}$ is the
single-particle Hamiltonian: \beq{P_0\equiv H =
\frac{\vec{p}^2}{2\mu} - \frac{Ze^2}{r},}\eeq $$ Z =\textrm{Nuclear
charge,}$$ $$\vec{r} = \textrm{relative coordinates,}$$ $$\mu =
\textrm{reduced mass.}$$

For this paper, the Hopf algebra of interest is the group algebra
$\mathbb{C}(SU(2)\times\mathbb{R})$ where $\mathbb{R}$ is time
translation, along with the coproduct $\Delta_{\chi\vec{n}}$. We are
interested in its concrete realization, denoted here as
$H_{\chi}(SU(2)\times\mathbb{R})$, on multi-electron states. We now
describe a convenient basis for this Hilbert space and evaluate the
coproducts $\Delta_{\chi\vec{n}}(H)$ and $\Delta_{\chi\vec{n}}(J_i)$
of the Hamiltonian and angular momentum in this basis.

The single particle basis we choose consists of eigenstates of $H$
and is
\beq{|N,~l\rangle\otimes|\alpha\rangle_{\vec{n}}\equiv|N,~l,~\alpha\rangle_{\vec{n}},~~~\alpha
= \pm 1 }\eeq where $N$ and $l$ are the principal quantum number and
orbital angular momentum and $|\alpha\rangle_{\vec{n}}$ denotes the
eigenstates of $\vec{\sigma}\cdot\vec{n}$ ($\sigma_i$ being Pauli
matrices) with eigenvalues $\alpha$:
\beq{H|N,~l,~\alpha\rangle_{\vec{n}} =
E_N|N,~l,~\alpha\rangle_{\vec{n}},}\eeq $$E_N = -\frac{Z\times
13.6}{N^2}\textrm{eV} = \textrm{energy for principal quantum
number}~N $$ \beq{\vec{\sigma}\cdot\vec{n}
|N,~l,~\alpha\rangle_{\vec{n}} = \alpha
|N,~l,~\alpha\rangle_{\vec{n}}.}\eeq

 The state vector $|N,~l,~\alpha\rangle_{\vec{n}}$ is
$|N,~l\rangle\otimes|\alpha\rangle_{\vec{n}}$ where the spin vector
$|\alpha\rangle_{\vec{n}}$ can be constructed as follows. Let
$g(\vec{n})\in SU(2)$ (in its defining representation) such
that~\cite{ClassTop, Group} \beq{g(\vec{n})\sigma_3 g(\vec{n})^\dag
= \vec{\sigma}\cdot\vec{n}}\eeq and let \beq{\sigma_3
|\alpha\rangle_{\hat{k}} = \alpha|\alpha\rangle_{\hat{k}},~~\hat{k}
= \left(\begin{array}{ccc} 0, & 0, & 1 \end{array}\right)}\eeq so
that \beq{|+\rangle_{\hat{k}} = \left(\begin{array}{c} 1\\0
\end{array}\right), ~~ |-\rangle_{\hat{k}} = \left(\begin{array}{c} 0\\1
\end{array}\right).}\eeq Then \beq\label{grot}{g(\vec{n})
|\alpha\rangle_{\hat{k}} = |\alpha\rangle_{\vec{n}}~.}\eeq Note that
$g(\vec{n})$ is not unique as both $g(\vec{n})$ and
$g(\vec{n})e^{i\sigma_3\theta}$ rotate $\sigma_3$ to
$\vec{\sigma}\cdot\vec{n}$. This ambiguity will disappear when we
compute rates. We also do not need an explicit choice of
$g(\vec{n})$ to calculate rates.

Next we calculate $\Delta_{\chi\vec{n}}(H)$ and
$\Delta_{\chi\vec{n}}(\vec{J})$.

As for $\Delta_{\chi\vec{n}}(H)$ and
$\Delta_{\chi\vec{n}}(\vec{n}\cdot\vec{J})$, they are not affected by
$\chi$ since $H$ and $\vec{n}\cdot\vec{J}$ commute and $G_{\chi\vec{n}}$
contains only these operators. (Hereafter $G_{\chi\vec{n}}$ denotes
Eq.(\ref{gtwist}) on the electronic states with spin included.) Thus
\beq\label{coH}{\Delta_{\chi\vec{n}}(H) = H\otimes 1 + 1\otimes
H,}\eeq \beq\label{conJ}{\Delta_{\chi\vec{n}}(\vec{n}\cdot\vec{J}) =
\vec{n}\cdot\vec{J}\otimes 1 + 1\otimes \vec{n}\cdot\vec{J}.}\eeq

 The coproduct for the remaining components of $\vec{J}$ can be
evaluated as follows. Let $\vec{n}^a$, $(a=1,2)$, $\vec{n}$ be an
orthonormal positively oriented coordinate system so that
$\vec{n}^1\wedge\vec{n}^2= \vec{n}$, and let
$$\vec{n}^{\pm}\cdot\vec{J}= (\vec{n}^1 \pm i\vec{n}^2)\cdot\vec{J}. $$ Then
\beq{\left[ \vec{n}\cdot\vec{J},~ \vec{n}^{(\pm)}\cdot\vec{J}\right] = \pm
\vec{n}^{(\pm)}\cdot\vec{J},}\eeq \beq{\left[ \vec{n}^{(+)}\cdot\vec{J},~
\vec{n}^{(-)}\cdot\vec{J}\right] = 2~\vec{n}\cdot\vec{J}.}\eeq From, this it
follows that
\beq\label{conpmJ}{\Delta_{\chi\vec{n}}(\vec{n}^{(\pm)}\cdot\vec{J}) =
\vec{n}^{(\pm)}\cdot\vec{J}\otimes e^{\mp\frac{i}{2}\chi P_0} +
e^{\pm\frac{i}{2}\chi P_0}\otimes\vec{n}^{(\pm)}\cdot\vec{J}.}\eeq

\section{The Electronic States of B\lowercase{e}}

The nucleus of Be has $Z=4$. We put two of the four electrons of
neutral Be in the $N=1$ level. The remaining two are put in the
$N=2,l=0$ level. The choice $l=0$ for all these levels is deliberate
as it greatly simplifies the calculations.

The equations Eq.(\ref{coH}), Eq.(\ref{conJ}) show that energy and
$\vec{n}\cdot\vec{J}$ are additive in the twist antisymmetrized levels
$\frac{1-\tau_{\chi\vec{n}}}{2}\left(|N,~l,~\alpha\rangle_{\vec{n}}\otimes
|N',~l',~\alpha'\rangle_{\vec{n}}\right)$. We have
\beq\label{actionofcoH}{\Delta_{\chi\vec{n}}(H)
\frac{1-\tau_{\chi\vec{n}}}{2} |N,~0,~\alpha\rangle_{\vec{n}}\otimes
|N',~0,~\alpha'\rangle_{\vec{n}} = (E_N + E_{N'})
\frac{1-\tau_{\chi\vec{n}}}{2}|N,~0,~\alpha\rangle_{\vec{n}} \otimes
|N',~0,~\alpha'\rangle_{\vec{n}},}\eeq
\beq\label{actionofconJ}{\Delta_{\chi\vec{n}}(\vec{n}\cdot\vec{J})
\frac{1-\tau_{\chi\vec{n}}}{2} |N,~0,~\alpha\rangle_{\vec{n}}\otimes
|N',~0,~\alpha'\rangle_{\vec{n}} = \frac{1}{2}(\alpha + \alpha')
\frac{1-\tau_{\chi\vec{n}}}{2}|N,~0,~\alpha\rangle_{\vec{n}} \otimes
|N',~0,~\alpha'\rangle_{\vec{n}}.}\eeq As for
$\Delta_{\chi\vec{n}}(\vec{n}^{(\pm)}\cdot\vec{J})$, we find,
\beq\label{actionof+onm}{\Delta_{\chi\vec{n}}(\vec{n}^{(+)}\cdot\vec{J})\left\{\begin{array}{c}
|N,~0,~+1\rangle_{\vec{n}}\otimes |N',~0,~-1\rangle_{\vec{n}}, \\
|N,~0,~-1\rangle_{\vec{n}}\otimes
|N',~0,~+1\rangle_{\vec{n}}\end{array}\right\}
 = \left\{\begin{array}{c} e^{\frac{i}{2}\chi E_N}
|N,~0,~+1\rangle_{\vec{n}}\otimes |N',~0,~+1\rangle_{\vec{n}}, \\
e^{-\frac{i}{2}\chi E_{N'}} |N,~0,~+1\rangle_{\vec{n}}\otimes
|N',~0,~+1\rangle_{\vec{n}} \end{array}\right\};}\eeq
\beq\label{actionof+ons}{\Delta_{\chi\vec{n}}(\vec{n}^{(+)}\cdot\vec{J})\left\{\begin{array}{c}
|N,~0,~+1\rangle_{\vec{n}}\otimes |N',~0,~+1\rangle_{\vec{n}}, \\
|N,~0,~-1\rangle_{\vec{n}}\otimes |N',~0,~
-1\rangle_{\vec{n}}\end{array}\right\}
 = \left\{\begin{array}{c} 0, \\ \begin{split} & e^{-\frac{i}{2}\chi E_{N'}}
|N,~0,~+1\rangle_{\vec{n}}\otimes |N',~0,~-1\rangle_{\vec{n}} \\ & +
e^{\frac{i}{2}\chi E_N} |N,~0,~-1\rangle_{\vec{n}}\otimes
|N',~0,~+1\rangle_{\vec{n}}\end{split}\end{array} \right\};}\eeq
\beq\label{actionof-onm}{\Delta_{\chi\vec{n}}(\vec{n}^{(-)}\cdot\vec{J})\left\{\begin{array}{c}
|N,~0,~+1\rangle_{\vec{n}}\otimes |N',~0,~-1\rangle_{\vec{n}}, \\
|N,~0,~-1\rangle_{\vec{n}}\otimes |N',~0,~
+1\rangle_{\vec{n}}\end{array}\right\}  = \left\{\begin{array}{c}
e^{\frac{i}{2}\chi E_{N'}}
|N,~0,~-1\rangle_{\vec{n}}\otimes |N',~0,~-1\rangle_{\vec{n}}, \\
e^{-\frac{i}{2}\chi E_N} |N,~0,~-1\rangle_{\vec{n}}\otimes
|N',~0,~-1\rangle_{\vec{n}} \end{array}\right\};}\eeq
\beq\label{actionof-ons}{\Delta_{\chi\vec{n}}(\vec{n}^{(-)}\cdot\vec{J})\left\{\begin{array}{c}
|N,~0,~+1\rangle_{\vec{n}}\otimes |N',~0,~+1\rangle_{\vec{n}}, \\
|N,~0,~-1\rangle_{\vec{n}}\otimes |N',~0,~
-1\rangle_{\vec{n}}\end{array}\right\}  = \left\{\begin{array}{c}
\begin{split} & e^{\frac{i}{2}\chi E_{N'}} |N,~0,~-1\rangle_{\vec{n}}\otimes
|N',~0,~+1\rangle_{\vec{n}}\\ & + e^{-\frac{i}{2}\chi E_N}
|N,~0,~+1\rangle_{\vec{n}}\otimes |N',~0,~-1\rangle_{\vec{n}}, \end{split} \\
0
\end{array}\right\}.}\eeq

\subsection{\it \large{\large{The two-electron ground state}}}

For $\chi =0$, it is unique, being the (untwisted) spin-singlet
state, \beq{\frac{1-\tau_0}{\sqrt{2}}
|1,~0,~+1\rangle_{\vec{n}}\otimes |1,~0,~-1\rangle_{\vec{n}} =
\frac{1}{\sqrt{2}}\left(|1,~0,~+1\rangle_{\vec{n}}\otimes
|1,~0,~-1\rangle_{\vec{n}} - |1,~0,~-1\rangle_{\vec{n}}\otimes
|1,~0,~+1\rangle_{\vec{n}} \right) }\eeq with energy $2E_{10}$.

As $\chi$ is changed away from $0$, thus vector is deformed to
\beq{ \label{ground}\begin{split} |1~1\rangle_{\chi\vec{n}} & =
  \frac{1-\tau_{\chi\vec{n}}}{\sqrt{2}}
|1,~0,~+1\rangle_{\vec{n}}~|1,~0,~-1\rangle_{\vec{n}} \\  &
= \frac{1}{\sqrt{2}}\left[
|1,~0,~+1\rangle_{\vec{n}}~|1,~0,~-1\rangle_{\vec{n}} - e^{i\chi
E_1}|1,~0,~-1\rangle_{\vec{n}}~|1,~0,~+1\rangle_{\vec{n}}\right].\end{split}}\eeq
Its energy still remains $2E_{10}$ in view of
Eq.(\ref{actionofcoH}).

No new linearly independent state appears by continuity: if they had
appeared, then as $\chi\rightarrow 0$, the ground state would not be
unique. We can verify this assertion by calculating
$\frac{1-\tau_{\chi\vec{n}}}{\sqrt{2}}
|1,~0,~\alpha\rangle_{\vec{n}}\otimes
|1,~0,~\alpha'\rangle_{\vec{n}}$ for any choice of $\alpha$,
$\alpha'$ and verifying that it is either proportional to
Eq.(\ref{ground}) or zero.

The values of $\Delta_{\chi\vec{n}}(\vec{n}\cdot\vec{J})$ and
$\Delta_{\chi\vec{n}}(\vec{n}^{(\pm)}\cdot\vec{J})$ on
$|1~1\rangle_{\chi\vec{n}}$ are also zero from
Eq.(\ref{actionofconJ}), Eq.(\ref{actionof+onm}) and
Eq.(\ref{actionof-onm}). So it is a twisted spin-singlet with zero
(twisted) value for total angular momentum.

\subsection{\it \large{The two-electron excited state}}

The actual Pauli-forbidden transition we will calculate will use the
excited state \beq{\frac{1-\tau_{\chi\vec{n}}}{\sqrt{2}}
\left[|2,~0,~+1\rangle_{\vec{n}}~|3,~0,~+1\rangle_{\vec{n}}\right]
}\eeq which is part of a (twisted!) spin triplet with orbital
angular momentum $0$ and energy $E_2 + E_3$.

 For completeness, we here list all the spin triplet and singlet
components of the states with energy $E_2 + E_3$.

{\it \large{The triplet vectors}}

\beq\label{tripl1}{ \Delta_{\chi\vec{n}}(\vec{n}\cdot\vec{J}) =1 :
\frac{1}{\sqrt{2}}\left[
|2,~0,~+1\rangle_{\vec{n}}~|3,~0,~+1\rangle_{\vec{n}} -
e^{\frac{i}{2}\chi(E_3-E_2)}|3,~0,~+1\rangle_{\vec{n}}~|2,~0,~+1\rangle_{\vec{n}}\right].}\eeq
\beq{ \Delta_{\chi\vec{n}}(\vec{n}\cdot\vec{J}) =0 :
\frac{1}{2}\left[\begin{split} &
e^{\frac{i}{2}\chi E_3}|2,~0,~-1\rangle_{\vec{n}}~|3,~0,~+1\rangle_{\vec{n}}
-
e^{-\frac{i}{2}\chi E_2}|3,~0,~+1\rangle_{\vec{n}}~|2,~0,~-1\rangle_{\vec{n}}\\
 & +
e^{-\frac{i}{2}\chi
E_2}|2,~0,~+1\rangle_{\vec{n}}~|3,~0,~-1\rangle_{\vec{n}} -
e^{\frac{i}{2}\chi
E_3}|3,~0,~-1\rangle_{\vec{n}}~|2,~0,~+1\rangle_{\vec{n}}\end{split}\right].}\eeq
\beq{ \Delta_{\chi\vec{n}}(\vec{n}\cdot\vec{J}) =-1 :
\frac{1}{\sqrt{2}}\left[
|2,~0,~-1\rangle_{\vec{n}}~|3,~0,~-1\rangle_{\vec{n}} -
e^{-\frac{i}{2}\chi(E_3-E_2)}|3,~0,~-1\rangle_{\vec{n}}~|2,~0,~-1\rangle_{\vec{n}}\right].}\eeq

{\it \large{The singlet vector}}

\beq{ \Delta_{\chi\vec{n}}(\vec{n}\cdot\vec{J})
=\Delta_{\chi\vec{n}}(\vec{n}^{(\pm)}\cdot\vec{J})=0 :
\frac{1}{2}\left[\begin{split} &
e^{\frac{i}{2}\chi E_3}|2,~0,~-1\rangle_{\vec{n}}~|3,~0,~+1\rangle_{\vec{n}}
-
e^{-\frac{i}{2}\chi E_2}|3,~0,~+1\rangle_{\vec{n}}~|2,~0,~-1\rangle_{\vec{n}}\\
 & -
e^{-\frac{i}{2}\chi
E_2}|2,~0,~+1\rangle_{\vec{n}}~|3,~0,~-1\rangle_{\vec{n}} +
e^{\frac{i}{2}\chi
E_3}|3,~0,~-1\rangle_{\vec{n}}~|2,~0,~+1\rangle_{\vec{n}}\end{split}\right].}\eeq

In the above equations for the triplet and singlet states, the values of the $\vec{n}\cdot\vec{J}$
components of the angular momenta are specified next to each of the states by specifying the
values of $\Delta_{\chi\vec{n}}(\vec{n}\cdot\vec{J})$ on each of these states. This is similar to specifying the
values of the third component of angular momentum.

\section{The Non-Pauli Rate}

This section contains the formula Eq.(\ref{rate}) for confrontation
with experiments. The rest of this section is a derivation of this
formula.

\subsection{\it \large{Spin overlaps}}

The basic transition we focus on is from the triplet excited state
Eq.(\ref{tripl1}) for twist $G_{\chi\vec{n}}$ to the ground state
levels for twist $G_{\chi\vec{m}}$. That involves the calculation of
the overlap $_{\vec{m}}\langle\alpha'|\alpha\rangle_{\vec{n}}$ which
follows from Eq.(\ref{grot}):
\beq{_{\vec{m}}\langle\alpha'|\alpha\rangle_{\vec{n}} =
\left(g(\vec{m})^{\dag}g(\vec{n})\right)_{\alpha'\alpha}.}\eeq This
expression depends on the choice of $g(\vec{n})$, $g(\vec{m})$. But
in rates, we get its squared modulus. That depends only on
$\vec{m}\cdot\vec{n}$:
\beq\label{thedot}{|_{\vec{m}}\langle\alpha'|\alpha\rangle_{\vec{n}}|^2
= \frac{1}{2}\left[1 +
(-1)^{\frac{(\alpha'-\alpha)}{2}}\vec{m}\cdot\vec{n}\right].}\eeq

Here is a simple proof of Eq.(\ref{thedot}). The R.H.S is
$$g(\vec{m})^{\dag}_{\alpha'\rho}g(\vec{n})_{\rho\alpha}g(\vec{n})^{\dag}_{\alpha\lambda}g(\vec{m})_{\lambda\alpha'}$$
for fixed $\alpha$, $\alpha'$ and summed $\rho$, $\lambda$. Consider
$\alpha=\alpha'=1$: \beq{\begin{split}|_{\vec{m}}\langle
+1|+1\rangle_{\vec{n}}|^2 & =
Tr\left(g(\vec{n})\frac{1+\tau_3}{2}g(\vec{n})^{\dag}\right)\left(g(\vec{m})\frac{1+\tau_3}{2}g(\vec{m})^{\dag}\right)
\\ \nonumber & = \frac{1}{4}
Tr\left[1+\vec{n}\cdot\vec{\tau}\right]\left[1+\vec{m}\cdot\vec{\tau}\right]
\\ \nonumber & =
\frac{1}{2}\left[1+\vec{m}\cdot\vec{n}\right]\end{split}.}\eeq In a
similar way we can establish Eq.(\ref{thedot}) for any $\alpha$,
$\alpha'$.

We can now see the root of the non-Pauli transition. consider the
{\it twist symmetrized} ground state for twist along $\vec{m}$:
\beq{\frac{1+\tau_{\chi\vec{m}}}{\sqrt{2}}
|1,~0,~\alpha\rangle_{\vec{m}}~|1,~0,~\beta\rangle_{\vec{m}} =
\frac{1}{\sqrt{2}}\left[
|1,~0,~\alpha\rangle_{\vec{m}}~|1,~0,~\beta\rangle_{\vec{m}} +
e^{\frac{i}{2}\chi
E_1(\alpha-\beta)}|1,~0,~\beta\rangle_{\vec{m}}~|1,~0,~\alpha\rangle_{\vec{m}}\right].}\eeq
It is part of the spin triplet which with the twist antisymmetrized
singlet gives the four two-electron ground states.

The normalized radial wave function for principal quantum number $N$
can be denoted by $|N\rangle$. It is independent of the twist
direction. The tensor product $|N\rangle\otimes |M\rangle$ can then
be written as $|N,M\rangle$.

Now a generic perturbation, call it $V_0$, will have a non-zero
radial matrix element $\langle 1~1|V_0|2~3\rangle$ where $V_0$ is
regarded as spin-independent for illustration. Then the
Pauli-forbidden amplitudes are roughly proportional to this factor
multiplied by spin overlaps
$_{\vec{m}}\langle\alpha~\beta|\frac{(1+\tau_{\chi\vec{m}})}{\sqrt{2}}|+1~+1\rangle_{\vec{n}}$:
the spin-statistics connection does not permit
$\frac{(1+\tau_{\chi\vec{m}})}{\sqrt{2}}|\alpha~\beta\rangle_{\vec{m}}$.
But we will see that these overlaps are not zero. So there are
Pauli-forbidden transitions.

For $\chi=0$, let $V_0$ be a generic spin-independent perturbation
of the two-electron Hamiltonian. We do not show its dependence on
electron coordinates, but we can assume it to be symmetric in them
as it preserves statistics: \beq{[V_0, ~ \tau_0] = 0.}\eeq For
$\chi\neq 0$, we have to modify $V_0$ to $V_{\chi\vec{n}}$:
\beq\label{twistpot}{V_{\chi\vec{n}} = \frac{1}{2}\left[V_0 +
\tau_{\chi\vec{n}}V_0\tau_{\chi\vec{n}}\right]}\eeq so that it
preserves the twisted statistics. As $V_0$ is an external
perturbation which causes transitions between levels, it can be
time-dependent. The perturbation has additional time dependence as
$\vec{n}$ changes with time.

The perturbed two-electron Hamiltonian is \beq{H' =
\Delta_{\chi\vec{n}}(H) + V_{\chi\vec{n}}.}\eeq

 Let $\vec{\rho}(t)$ be a time-dependent unit vector which at $t=t_i$
is $\vec{n}$ and at time $t=t_f$ is $\vec{m}$. To leading order in
$V_{\chi\vec{n}}$, the transition matrix element from an initial
state $|I\rangle$ of energy $E_I$ at time $t_i$ to an orthogonal
final state $|F\rangle$ of energy $E_F$ at time $t_f$ is
$$-ie^{-i(t_f-t_i)E_f}\langle F|\int_{t_i}^{t_f}d\tau e^{i\tau
H}V_{\chi\vec{\rho}(\tau)}e^{-i\tau H} |I\rangle.$$ For us
\beq{|I\rangle = \frac{1-\tau_{\chi\vec{n}}}{\sqrt{2}}
|2,~0,~+1\rangle_{\vec{n}}~|3,~0,~+1\rangle_{\vec{n}},}\eeq with
$E_I = E_2+E_3$.

For $|F\rangle$, we choose a {\it Pauli-forbidden} ground state
$$\frac{1+\tau_{\chi\vec{m}}}{\sqrt{2}}|1,~0,~\alpha\rangle_{\vec{m}}~|1,~0,~\alpha'\rangle_{\vec{m}}$$
(This vector is not normalized if $\alpha=\alpha'$. We will fix that
problem later.)

From Eq.(\ref{twistpot}), we can see that $V_{\chi\vec{n}} = V_0 +
O(\chi)$. The explicit calculations below show that the amplitude is
$O(\chi)$ if $V_{\chi\vec{n}}$ is approximated by $V_0$. So we
approximate $V_{\chi\vec{n}}$ by $V_0$ in Eq.(\ref{twistpot})
neglecting terms of $O(\chi)$.

As $V_0$ is symmetric in electron coordinates, for the radial matrix
element, $\langle 1~1|V_0|2,~3\rangle = \langle
1~1|V_0|3,~2\rangle$.

We now use this identity to simplify the probability for transition
$P_{\chi}$ to any Pauli-forbidden ground state. That is obtained
from modulus squared of the amplitude by summing over $|F\rangle$
after normalizing them. But the projector to the Pauli-forbidden
ground states is \beq{ Q = |1,~1\rangle\langle
1,~1|\mathbb{I}_{\textrm{spin}} -
|1,~1\rangle_{\chi\vec{m}~\chi\vec{m}}\langle 1,~1|}\eeq where
$\mathbb{I}_{\textrm{spin}}$ is the unit operator on spin space.

Thus the probability of interest is \beq{P_{\chi} = \langle
I|\left(\int_{t_i}^{t_f}d\tau e^{i\tau
2E_1}V_0(\tau)e^{-i\tau(E_2+E_3)}\right)^*Q\left(\int_{t_i}^{t_f}e^{i\tau2E_1}V_0(\tau)e^{-i\tau(E_2+E_3)}\right)|I\rangle.}\eeq
This simplifies to the following on using the symmetry of $V_0$:
\beq{P_{\chi} = |\langle 1~1|\int_{t_i}^{t_f}d\tau e^{i\tau 2
E_1}V_0(\tau)e^{-i\tau(E_2+E_3)}|2~3\rangle|^2\times
P^{\chi}_{\textrm{SPIN}}}\eeq where \beq{ P^{\chi}_{\textrm{SPIN}} =
\frac{1}{2}|\left(1 - e^{\frac{i}{2}\chi(E_3-E_2)}\right)|^2\left[1
- \frac{1}{2}|\left(_{\vec{m}}\langle +~-| - e^{-i\chi
E_1}~_{\vec{m}}\langle -~+|\right) |+~+\rangle_{\vec{n}}|^2\right].
}\eeq As claimed, $P_{\chi}$ is $O(\chi^2)$.

$P^{\chi}_{\textrm{SPIN}}$ can be evaluated using Eq.(\ref{thedot}).
The result is \beq{P^{\chi}_{\textrm{SPIN}} =
2\sin^2(\frac{\chi}{4}\Delta
E)\left[1-\frac{1}{4}(1-(\vec{m}\cdot\vec{n})^2)(1-\cos(\chi
E_1))\right]}\eeq where $\Delta E = E_3-E_2$.

Here since $\vec{n}$ and $\vec{m}$ vary, it is best to average over
them using the rotationally invariant measure. We first average over
$\vec{m}$ by integrating over its polar and azimuthal angles
$\theta_m$, $\phi_m$ using the standard measure
$$\frac{d\omega_m}{4\pi},~~ d\omega_m=d\cos\theta_md\phi_m.$$ Then
\beq{\int\frac{d\omega_m}{4\pi}\mathbb{I} = 1,~
\int\frac{d\omega_m}{4\pi}m_i = 0,~ \int\frac{d\omega_m}{4\pi}m_im_j
= \frac{1}{3}\delta_{ij}}\eeq giving for the average $\langle
P_{\chi}\rangle$ of $P_{\chi}$, \beq{ \langle P_{\chi}\rangle =
\left\{|\langle 1~1|\int_{t_i}^{t_f}e^{i\tau
2E_1}V_0(\tau)e^{-i\tau(E_2+E_3)}|2~3\rangle|^2\right\}\times
\left\{\frac{1}{3}\left(5+\cos(\chi
E_1)\right)\sin^2(\frac{\chi}{4}\Delta E)\right\}.}\eeq There is no
need to average over $\vec{n}$ as this is $\vec{n}$-independent.

The magnitude of the prefactor in braces is that of a typical
probability for a Pauli-allowed process. Thus the branching ratio of
a Pauli-forbidden to a Pauli-allowed process is
\beq\label{rate}{B_{\chi} = \frac{1}{3}\left(5+\cos(\chi
E_1)\right)\sin^2(\frac{\chi}{4}\Delta E),~~\Delta E =E_3 -
E_2.}\eeq It is independent of $t_i$, $t_f$. It is this expression
we use to confront experiments as it is a ratio and may not be
sensitive to the details of its derivation.

\section{Experiments and Bounds on $\chi$}

The experiments searching for Pauli-forbidden transitions can be
broadly classified into atomic and nuclear experiments.  Here we
discuss each experiment separately.

Some of the above experiments give only lifetimes for the forbidden
processes. To obtain the branching ratios in such cases we multiply
the given rate with the typical lifetimes for such  processes. In
the case of an atomic process, we use the number $10^{-16}$ seconds
and for a nuclear process we use $10^{-23}$ seconds for typical
lifetimes.

\subsection*{Bounds from The Borexino Experiment}
 The Borexino collaboration has used its counting test facility
to obtain limits on the violation of the Pauli exclusion principle
(PEP) using nuclear transitions in $^{12}C$ and $^{16}O$ nuclei. The
method is to search for $\gamma$, $n$, $p$ and/or $\alpha$ emitted
in a non-Paulian transition of $1P$ shell nucleons to the filled
$1S_{1/2}$ shell in nuclei. Various stringent bounds were obtained
as a result.

We use the following result from the Borexino
experiment~\cite{Borexino}: \beq{\tau(^{12}C \rightarrow
^{12}\widetilde{C} + \gamma) \geq 5.0\times
10^{31}\textrm{years}.}\eeq In the above process,
$^{12}\widetilde{C}$ denotes an anomalous carbon nucleus with an
extra nucleon in the filled $K$ shell of $^{12}C$. This corresponds
to a branching ratio of the order of $10^{-62}$. We take $\Delta E$
for this process to be of the order of $1 \textrm{MeV}$ to get a
bound on $\chi$.

\subsection*{Bounds from The Kamiokande Detector}
 In this experiment searches were made for forbidden transitions in
$^{16}O$ nuclei and they obtain a bound on the ratio of forbidden
transitions to normal transitions. The bound for this ratio is
$< 2.3 \times 10^{-57}$~\cite{SKamiokande}. Again for this process
$\Delta E$ is assumed to be of the order of $1 \textrm{MeV}$.

\subsection*{Bounds from The NEMO Experiment}
 Similar to nucleon transitions, experiments searching for Pauli-forbidden
atomic transitions have also been performed. The NEMO
collaboration~\cite{NEMO} searches for anomalous
$^{12}\widehat{C}$ atoms which are those with $3$ $K$-shell
electrons. The method used is the $\gamma$ ray activation analysis
in a sample of boron where the impurity carbon has been removed
radiochemically. The bound on the existence of such atoms is given
by the ratio of abundances of $^{12}\widehat{C}$ to $^{12}C$: it is
$<2.5\times 10^{-12}$. It corresponds to a limit on the lifetime
with respect to violation of the Pauli principle by electrons in a
carbon atom of $\tau\geq 2\times 10^{21}\textrm{years}$. We take
$\Delta E$ for this process to be $272~\textrm{eV}$ to calculate a
bound on $\chi$.

The NEMO-2 collaboration has also performed nucleon transition
experiments~\cite{NEMO-2} and the limit obtained is \beq{\tau(^{12}C
\rightarrow ^{12}\widetilde{C} + \gamma) \geq 4.2\times
10^{24}\textrm{years}.}\eeq This corresponds to a branching ratio of
the order $<10^{-55}$ if we assume $\Delta E$ for this process to be
of the order of $1 \textrm{MeV}$.

\subsection*{Bounds from experiments at Maryland}
 Atomic transition experiments have been conducted by Ramberg and Snow
in Maryland using copper (Cu) atoms. The idea here is to introduce
new electrons into a copper strip and to look for the K X-rays that
would be emitted if one of these electrons were to be captured by a
Cu atom and cascade down to the $1S$ state despite the fact that the
$1S$ level was already filled with two electrons. The probability
for this to occur was found to be less than $1.76\times
10^{-26}$~\cite{Ramberg}. This corresponds to a lifetime of $\tau >
8.36\times 10^{3}\textrm{years}$. We assume $\Delta E$ for this
process to be of the order of $1.5\textrm{KeV}$.

\subsection*{Bounds from the VIP experiment}
 An improved version of the experiment at Maryland has been
performed by the VIP collaboration~\cite{VIP}. They improved the
limit obtained by Ramberg and Snow at Maryland by a factor of about
$40$. The limit on the probability of PEP violating interactions
between external electrons and copper is found to be less than
$4.5\times 10^{-28}$. Here again we take $\Delta E$ to be of the
order of $1.5 \textrm{KeV}$.

The bounds are summarized in Table (\ref{table1}).

\begin{table}
\begin{center}
\begin{tabular}{|c|c|c|c|}
\hline Experiment & Type & Bound on $\chi$ & Bound on $\chi$ \\
  &  & (Length scales) & (Energy scales) \\ \hline \hline
Borexino & Nuclear & $\lesssim 10^{-47}~\textrm{m}$ & $\gtrsim
10^{28}~\textrm{TeV}$  \\
Kamiokande & Nuclear & $10^{-42}~\textrm{m}$ &
$10^{23}~\textrm{TeV}$
\\ NEMO & Atomic & $10^{-12}~\textrm{m}$ & $10^{5}~\textrm{eV}$
\\ NEMO-2 & Nuclear & $10^{-41}~\textrm{m}$ & $10^{22}~\textrm{TeV}$
\\ Maryland & Atomic & $10^{-20}~\textrm{m}$ & $10~\textrm{TeV}$
\\ VIP & Atomic & $10^{-21}~\textrm{m}$ & $100~\textrm{TeV}$
\\
\hline
\end{tabular}
\end{center}
\caption{Bounds on the noncommutativity parameter $\chi$}
\label{table1}
\end{table}

\section{Time Quantization}

The algebra $B_{\chi\vec{n}}$ leads to time-quantization in units of
$\chi$ and therefore~\cite{Baltrgpaulo, Balpaulo} energy
nonconservation: it is conserved only mod $\frac{2\pi}{\chi}$. An
effect of this sort was first discovered by
Chaichian~\cite{Chaitime} for a cylindrical noncommutative
spacetime. Quantum physics on such spacetime including scattering
theory was later developed in~\cite{Balpaulo}.

Time quantization comes about as follows. From Eq.(\ref{Bplane}),
one sees that $x_0$ generates rotations around $\vec{n}$ and that
$e^{i\frac{2\pi}{\chi}x_0}$, being $2\pi$ rotation, acts as identity
on $x_i$. Being a time exponential, it also commutes with momentum
operators. Thus it is in the center of the algebra generated by
$B_{\chi\vec{n}}$ and by its momentum operators. Hence it is a
multiple of the identity in an irreducible representation of the
latter: \beq\label{time}{e^{i\frac{2\pi}{\chi}x_0} =
e^{i\phi}\mathbb{I},}\eeq  $e^{i\phi}$ being characteristic of the
representation.

A consequence of Eq.(\ref{time}) is that the spectrum
$\textrm{spec}~x_0$ of $x_0$ is quantized: \beq{\textrm{Spec}~ x_0 =
\chi\left(\mathbb{Z} + \frac{\phi}{2\pi}\right).}\eeq As explained
in~\cite{Balpaulo, Baltrgpaulo}, a quantum field $\psi$ is defined
only on the spectrum of time operator $x_0$. Time translations are
from one point of this spectrum to another, so that only the time
translations $$(e^{i\chi P_0})^N,~~N\in\mathbb{Z}$$ exist on quantum
fields.

But then $P_0$ and $P_0+\frac{2\pi}{\chi}M$, $M\in\mathbb{Z}$
generate the same time translation. Due to this we can anticipate
energy conservation only mod $\frac{2\pi}{\chi}$ in scattering
processes. This anticipation is correct. In~\cite{Baltrgpaulo},
scattering theory with time quantization has been developed and
energy is found to be conserved only mod $\frac{2\pi}{\chi}$.

An interesting application of such time quantization is to
extra-dimensional models. Thus for example if spacetime is
$M^4\times S^1$ where $M^4$ is our four-dimensional spacetime, and
the time operator $x_0$ fails to commute with the $e^{i\phi}$ which
generates the algebra of functions on $S^1$, \beq{x_0e^{i\phi} =
e^{i\phi}x_0 + \chi e^{i\phi},}\eeq then scattering theory on $M^4$
will conserve energy only mod $\chi$. No further interaction is
needed for this energy nonconservation to occur.

Such energy nonconservation can be tested by experiments.
Unfortunately, we know of no recent experiment to test energy
conservation.

\section{Final Remarks}

Non-Pauli transitions are expected to occur on the Moyal plane $\mathcal{A}_{\theta}$ as
well. But while the Moyal plane has the defining relations \beq\label{Moyal}{[x_{\mu},~x_{\nu}] = i\theta_{\mu\nu}}\eeq \beq{\theta_{\mu\nu}=-\theta_{\nu\mu}=\textrm{constant}}\eeq which are manifestly invariant under translations for the coordinates $x_{\mu}\rightarrow x_{\mu}+a_{\mu}$ ($a_{\mu}$=constant), they are not invariant under the naive rotation of coordinates. So the Moyal plane is not adapted to discuss atomic processes where rotational invariance plays a crucial role. That makes the calculations complicated. 



The Bose and Fermi sectors of a local quantum field theory are
superselected. But here we find transitions between these sectors.

There is no necessary contradiction due to the fact that the models
of this paper seem to violate a superselection rule which has been proved from general principles of quantum theory~\cite{WallyMessiah}. According to Greenberg and Messiah~\cite{WallyMessiah}, this rule does not require even the principles of local quantum field theories for its validity. The reason for this apparent violation is as follows. In the model considered here, the flip operator $\tau_{\chi\vec{n}}$ and hence what is meant by twisted Bose and Fermi particles itself changes with time. This situation does not occur in standard quantum physics and is in fact the source of non-Pauli effects. Also for fixed $\vec{n}$, matrix elements of observables between twisted Bose and twisted Fermi states are zero so that in this sense there is no violation of superselection rules.  

Note that the noncommutative models being discussed here are designed to probe energy
scales vastly higher than those where local quantum theories have
been tested. But certainly a deeper study of this violation is
important.

\section{Acknowledgements}
The authors are very grateful to Gianpiero Mangano, who helped us
with important suggestions at every stage of this work. We are also grateful
to the referee of our earlier paper~\cite{Pauli}, submitted to Physical Review
Letters, who greatly helped us with suggestions for improving that paper.
Thanks are due to Prof.T.R.Govindarajan for the wonderful hospitality at IMSc,
Chennai. This work was supported in part by DOE under the grant
number DE-FG02-85ER40231. The work of APB was also supported by the
Department of Science and Technology, India.

\end{document}